\documentclass{article}
\usepackage{amssymb,amsmath}
\usepackage{epsfig}
\usepackage{subfigure}
\newtheorem{rem}{Remark}[section]
\newcommand{\br}{\begin{rem}}
\newcommand{\er}{\end{rem}}
\newtheorem{ex}{Example}[section]
\newcommand{\bex}{\begin{ex}}
\newcommand{\eex}{\end{ex}}
\newtheorem{Def}{Definition}[section]
\newcommand{\bd}{\begin{Def}}
\newcommand{\ed}{\end{Def}}
\newtheorem{theorem}{Theorem}[section]
\newcommand{\bt}{\begin{theorem}}
\newcommand{\et}{\end{theorem}}
\newtheorem{lemma}{Lemma}[section]
\newcommand{\bl}{\begin{lemma}}
\newcommand{\el}{\end{lemma}}
\newcommand{\be}{\begin{equation}}
\newcommand{\ee}{\end{equation}}
\newcommand{\bea}{\begin{eqnarray}}
\newcommand{\eea}{\end{eqnarray}}
\newcommand{\pa}{\partial}
\newcommand{\nn}{\nonumber}
\newcommand{\adots}{\mathinner{\mkern2mu\raise1pt\hbox{.}\mkern2mu
\raise4pt\hbox{.}\mkern2mu\raise7pt\hbox{.}\mkern1mu}}

\title{A Kaluza-Klein Reduction of Super-integrable Systems}
\author{Allan P. Fordy,  \\
School of Mathematics, University of Leeds, \\
Leeds LS2 9JT, UK.\\
e-mail a.p.fordy@leeds.ac.uk}

\begin{document}

\maketitle

\bibliographystyle{plain}

\begin{abstract}
Given a super-integrable system in $n$ degrees of freedom, possessing an integral which is linear in momenta, we use the ``Kaluza-Klein construction" in reverse to reduce to a lower dimensional super-integrable system.  We give two examples of a reduction from 3 to 2 dimensions.  The constant curvature metric (associated with the kinetic energy) is the same in both cases, but with two different super-integrable extensions.  For these, we use different elements of the algebra of isometries of the kinetic energy to reduce to $2-$dimensions.  Remarkably, the isometries of the reduced space can be derived from those of the $3-$dimensional space, even though it requires the use of {\em quadratic} expressions in momenta.
\end{abstract}



\section{Introduction}

We first describe the standard Kaluza-Klein construction.  Suppose we have a Hamiltonian with ``electromagnetic terms'' in $n-$dimensional configuration space
\be\label{kk-general1}
H = \frac{1}{2}\, \sum_{i=1}^n g^{ij}(p_i-A_i)(p_j-A_j) + h,
\ee
where $g^{ij}, \, A_i,\, h$ are functions of ${\bf q}$ only.  The functions $g^{ij}$ represent the coefficients of the {\em inverse} of a metric $g_{ij}$ (not necessarily Riemannian).  The determinant of the matrix of these coefficients should, therefore, be non-zero.  The functions $A_i$ are interpreted as the components of the electromagnetic vector potential and $h$ as the potential of some external force.  It should be noted that {\em any} Hamiltonian of the form
$$
H = \frac{1}{2}\, \sum_{i=1}^n g^{ij} p_i p_j + \sum_{i=1}^n \tilde A_i p_i + \tilde h
$$
can be put in the form (\ref{kk-general1}) by ``completing the square''.  Notice that
$$
\{p_i-A_i,p_j-A_j\} = \pa_iA_j-\pa_jA_i,
$$
which are just the components of the electromagnetic field tensor $F_{ij}$.  When this vanishes, the electromagnetic field is {\em trivial} and we can make a gauge transformation (in the Lagrangian/Hamiltonian sense) to remove the the $A_i$ from the picture:
$$
\{p_i-A_i,p_j-A_j\}=0,\;\; \{q_i,p_j-A_j\}=\{q_i,p_j\}=\delta_{ij} \Rightarrow (q_i,\hat p_i) \;\;\;\mbox{are canonically conjugate variables},
$$
where $\hat p_i = p_i-A_i$.  Thus, in coordinates $(q_i,\hat p_i)$, the Hamiltonian (\ref{kk-general1}) reduces to the standard case with no electromagnetic terms.

This construction introduces an additional dimension, with additional coordinates $(q_{n+1},p_{n+1})$, so that $q_1,\dots ,q_{n+1},p_1,\dots ,p_{n+1}$ are canonical, and modifies the Hamiltonian (\ref{kk-general1}) in the following way
\be\label{kk-general2}
H = \frac{1}{2}\, \sum_{i=1}^n g^{ij}(p_i-A_i p_{n+1})(p_j-A_j p_{n+1}) + g^{n+1\, n+1} p_{n+1}^2 + \bar h,
\ee
where part of the original $h$ may have been absorbed into the $g^{n+1\, n+1}$ term.  In this way the vector potential in incorporated into a metric on a higher dimensional space.  The $g^{n+1\, n+1}$ term is chosen for convenience, but should at least make the new ``inverse metric'' nonsingular.  The purpose of Kaluza-Klein theory was to unify gravitation and electromagnetic theory by raising the dimension of space-time from 4 to 5 dimensions.  There was no interest in complete integrability.

However, it can be seen that, by construction, the new metric and the Hamiltonian do not depend upon the variable $q_{n+1}$, so $p_{n+1}$ is a simple first integral (the Noether constant corresponding to a translation in the $q_{n+1}$ direction).  This gives us the route back down to $n-$dimensions, by reducing to the invariant submanifold $p_{n+1}=b$ (a constant).  Our purpose in this paper is to use this mechanism to make {\em nontrivial} reductions of a given super-integrable system to one of lower dimension.  In fact, we only discuss the reduction from 3 degrees of freedom to two.  The general framework will be described in Section \ref{genkk} and applied to specific examples in Sections \ref{concurv-super1} and \ref{concurv-super2}.

We finish this introduction with a brief reminder of the definitions of complete and super-integrability.

A Hamiltonian system of $n$ degrees of freedom, Hamiltonian $H$, is said to be {\em completely integrable in the Liouville sense} if we have $n$ independent functions $I_n$, which are {\em in involution} (mutually Poisson commuting), with $H$ being a function of these and typically just one of them. Whilst $n$ is the maximal number of independent functions which can be {\em in involution}, it is possible to have further integrals of the Hamiltonian $H$, which necessarily generate a non-Abelian algebra of integrals of $H$.  The maximal number of additional {\em independent} integrals is $n-1$, since the ``level surface'' of $2n-1$ integrals (meaning the intersection of individual level surfaces) is just the (unparameterised) integral curve.  Well known elementary examples are the isotropic harmonic oscillator, the Kepler system and the Calogero-Moser system.

The quadratures of complete integrability are often achieved through the separation of variables of the Hamilton-Jacobi equation.  The solution of a maximally super-integrable system can also be calculated purely algebraically (albeit implicitly), requiring just the solution of the equations $I_k=c_k,\, k=1,\dots ,2n-1$.  Maximally superintegrable systems have a number of interesting properties: they can be separable in more than one coordinate system; all bounded orbits are closed; they give rise to interesting Poisson algebras with polynomial Poisson relations.  The idea can be extended to {\em quantum integrable systems}, with first integrals replaced by commuting differential operators. For some examples of superintegrable quantum systems it is possible to use the additional commuting operators to build sequences of eigenfunctions \cite{f07-1}.

There is a large literature on the classification and analysis of superintegrable systems (see the review \cite{13-2}) and they naturally occur in many applications in physics (additional integrals being referred to as ``hidden symmetries'' \cite{14-2}).

\section{The General Framework}\label{genkk}

Since the Kaluza-Klein reduction is concerned with (pseudo-)Riemannian metrics, we consider Hamiltonians in ``natural form'' (the sum of kinetic and potential energies).  After the reduction, the lower dimensional Hamiltonian will have electromagnetic terms, which could turn out to be ``trivial'' (with zero electromagnetic tensor).  In Sections \ref{concurv-super1} and \ref{concurv-super2} we will start with a {\em given} $3-$dimensional metric and reduce to $2-$dimensions, but here we first make a few general remarks.

For a manifold with coordinates $(q_1,\dots ,q_n)$, metric coefficients $g_{ij}$, with inverse $g^{ij}$, the geodesic equations are Hamiltonian, with {\em kinetic energy}
\be\label{Ham-h2}
H = \frac{1}{2}\, \sum_{i,j=1}^n g^{ij}p_ip_j,\quad\mbox{where}\quad  p_i=\sum_k g_{ik}\dot q_k.
\ee
For a metric with isometries, the infinitesimal generators (Killing vectors) give rise to first integrals, which are {\em linear} in momenta (Noether constants).
When the space is either flat or constant curvature, it possesses the maximal group
of isometries, which is of dimension $\frac{1}{2}n(n+1)$.  In this case, (\ref{Ham-h2}) is actually the second order {\em Casimir} function of the symmetry algebra (see \cite{74-7}).

Starting with a kinetic energy $H_0$ corresponding to a flat or constant curvature metric, we can use its isometry algebra to build commuting quadratic functions.  Indeed, in flat and constant curvature spaces {\em all} homogeneously quadratic integrals are just quadratic forms of the isometries.  We can then add potential functions.
In 3 degrees of freedom, this means that we would have the kinetic energy $H_0$, together with 2 homogeneously quadratic integrals of $H_0$ (corresponding to Killing tensors) and 3 functions of just the coordinates $q_i$:
$$
H = H_0 + h({\bf q}), \;\;  F_1=F^{(2)}_1 + F^{(0)}_1 , \;\;  F_2=F^{(2)}_2 + F^{(0)}_2, \;\;\;\mbox{satisfying}\;\;\; \{H,F_1\} = \{H,F_2\} = \{F_1,F_2\} = 0.
$$
Since we start with a constant (possibly zero) curvature metric, we build the quadratic parts $F^{(2)}_i$ from quadratic forms of the Killing vectors (Noether constants).  The Poisson relations of the symmetry algebra help us choose specific quadratic pairs which satisfy $\{F_1,F_2\} = 0$.

The choice of quadratic integrals means that our systems will be separable.  The calculation of separable potentials is standard and it is well known that in the standard orthogonal coordinate systems, with separable kinetic energies, we can add potentials which depend upon a number of arbitrary functions of a single variable \cite{76-8}.  If a complete (possessing $n$ parameters) solution of the Hamilton-Jacobi equation is found, then, by Jacobi's theorem, these parameters, when written in terms of the canonical variables, are quadratic first integrals of $H$.  The problem has also been posed in the ``opposite'' direction: given a pair of Poisson commuting, homogeneously quadratic integrals (in two degrees of freedom) what sort of potentials can be added, whilst maintaining commutativity?  This is a classical problem (see Whittaker \cite{88-4}, chapter 12, section 152) and leads to the Bertrand-Darboux equation for the potential \cite{88-8,08-7}.

To build a super-integrable system, we can start with an involutive system $H, F_1, F_2$, as above, and add two more functions $F_3,F_4$, satisfying $\{H,F_3\}=\{H,F_4\}=0$, but with the (now) {\em given} $H$.  The functions should be chosen to be {\em functionally independent}, so the Jacobian matrix
$$
\frac{\pa(H,F_i)}{\pa {\bf x}},\quad\mbox{where}\quad   {\bf x} = (q_1,\dots , p_3),
$$
has \underline{maximal rank} $5$, since in this case, the level surface ${\cal S} = \{{\bf x}: H=c_0,F_i=c_i\}_{i=1}^4$, has dimension \underline{one}, so represents an (unparameterised) trajectory of the dynamical system.

Demanding the existence of these additional integrals reduces the arbitrary functions of the separable potential to specific functions which depend upon only a {\em finite number of parameters}.

In \cite{f17-5} we used the above approach to construct a number of super-integrable systems, including the examples of Sections \ref{concurv-super1} and \ref{concurv-super2}.  The more usual way is to consider functions with general quadratic parts and to use commutativity to fix parameters (see \cite{13-2}), but, by using the algebra of infinitesimal isometries of the metric, we can exploit the discrete automorphisms of the algebra, which induce automorphisms of the Poisson algebra of our super-integrable system.  This will be seen in the examples of Sections \ref{concurv-super1} and \ref{concurv-super2}.

\subsection{The Kaluza-Klein Reduction}\label{kk-reduce-gen}

For our Kaluza-Klein reduction, the given super-integrable system must have at least one integral, $I$, which is {\em first order} in momenta.  This must necessarily Poisson commute with {\em both} the kinetic and potential energies, so is an element of the isometry algebra of the metric.  We can always ``adapt coordinates'' to this geometric symmetry, which in the Poisson context, means that we choose coordinates so that $I = P_3$ (we are describing the process in the case of 3 degrees of freedom).  This will render a Hamiltonian which is independent of the coordinate $Q_3$, so we can consider the reduction to the space with (spatial) coordinates $Q_1,Q_2$, but at this point we can say nothing about the integrability (or otherwise) of the reduced system.

There is a lot of freedom in choosing the coordinates $Q_1,Q_2$, such that $\{Q_i(q_1,q_2,q_3),I\}=0$, so we can stipulate that these are orthogonal with respect to the $3-$dimensional metric $g$, so that the $2-$dimensional reduced metric is diagonal.  At this point we could just calculate the first order invariants of the new kinetic energy (corresponding to Killing vectors). Since the reduced metric is again constant curvature, it will have 3 Killing vectors.  However, it is possible to {\em construct} these as reductions of invariants of the $3-$dimensional kinetic energy.  In the examples below, one of these is just an ``old'' first order invariant written in the new coordinates, but the others are less obvious, being derived from certain {\em quadratic} expressions.  For our reduced system to be at least integrable (and preferably super-integrable) we need to find enough integrals of the $3-$dimensional system which {\em Poisson commute} with the geometric symmetry $I$.  In this case, these integrals will be independent of the new $Q_3$, so can be reduced to the $2-$ dimensional space.

Whilst this is certainly not the generic situation, such examples are quite common.

\subsection{A Constant Curvature Metric and its Isometries}\label{3dmetric}

The two examples of Sections \ref{concurv-super1} and \ref{concurv-super2} are derived from Hamiltonians with the same {kinetic energy} (and therefore $3-$dimensional metric) but have different potentials and Poisson algebras.  The reductions are with respect to different geometric symmetries, so the reduced metrics are different.  The $3-$dimensional metric, with its curvature properties and isometries are given below.

Consider the Hamiltonian (kinetic energy)
\be\label{H12}
H_0 =  q_1^2(p_1^2-p_2^2-p_3^2),\quad \mbox{with}\quad g^{ij} = {\rm{diag}}\left(q_1^2,-q_1^2,-q_1^2 \right),
\ee
which has constant curvature, with isometry algebra (see \cite{f17-5})
\bea
&& e_1 = p_2, \quad h_1 = -2(q_1p_1+q_2p_2+q_3p_3), \quad  f_1 = -2q_1q_2p_1+(q_3^2-q_1^2-q_2^2)p_2-2q_2q_3p_3 ,\nn\\[-2mm]
&&  \label{diag-alg}\\[-2mm]
&& e_2 = p_3, \quad h_2 = 2(q_2p_3-q_3p_2), \quad  f_2 = -4q_3(q_1p_1+q_2p_2)-2(q_1^2-q_2^2+q_3^2)p_3 , \nn
\eea
with Poisson relations given in Table \ref{Tab:6Symm_alg}.  We see that the algebra has rank 2.  In fact, it is easy to see that this is isomorphic to $\mathbf{so}(1,3)$.
\begin{table}[h]\centering
\caption{The $6-$dimensional isometry algebra of (\ref{H12})}\label{Tab:6Symm_alg}\vspace{3mm}
\begin{tabular}{|c||c|c|c||c|c|c|}
\hline   &$e_1$    &$h_1$     &$f_1$   &$e_2$   &$h_2$    &$f_2$    \\[.10cm]\hline\hline
$e_1$    &0        &$2e_1$    &$-h_1$  &0       &$-2e_2$    &$-2h_2$      \\[1mm]\hline
$h_1$    &$-2e_1$   &0        &$2f_1$  &$-2e_2$ &0        &$2f_2$   \\[1mm]\hline
$f_1$    &$h_1$    &$-2f_1$   &0       &$-h_2$  &$-f_2$    &0        \\[1mm]\hline\hline
$e_2$    &0         &$2e_2$   &$h_2$ &0       & $2e_1$   & $-2h_1$   \\[1mm]\hline
$h_2$    &$2e_2$    &0        &$f_2$  &$-2 e_1$   &0        &$-4 f_1$    \\[1mm]\hline
$f_2$    &$2h_2$   &$-2f_2$  &0       &$2 h_1$   &$4 f_1$   &0        \\[1mm]\hline
\end{tabular}
\end{table}

The quadratic Casimir of this algebra is proportional to $H_0$:
\begin{subequations}
\be\label{H03d}
H_0 = e_1f_1+\frac{1}{4} h_1^2+ \frac{1}{4} (2e_2 f_2-h_2^2).
\ee
There is a second independent (fourth order) Casimir element of the abstract algebra, but in this representation it is a perfect square and {\em zero}, giving the quadratic constraint
\be\label{cas2}
e_1 f_2+h_1 h_2-2 f_1 e_2=0.
\ee
\end{subequations}
This is the only constraint, since the Jacobian of the six isometries has rank 5.

\br[Maximally Super-Integrable]
The geodesic equations corresponding to (\ref{H12}) have 5 functionally independent first integrals, so are themselves a maximally super-integrable system.
\er

This algebra has the following useful pair of involutive automorphisms:
\bea
 \iota_1:\;  (e_1,h_1,f_1,e_2,h_2,f_2)  &\mapsto &   \left(f_1,-h_1,e_1,-\frac{1}{2} f_2,-h_2,-2 e_2\right),  \nn\\[-2mm]
     &&    \label{i123}       \\[-2mm]
 \iota_{23}:\; (e_1,h_1,f_1,e_2,h_2,f_2) &\mapsto & \left(e_2,h_1,\frac{1}{2} f_2,e_1,-h_2,2 f_1\right).  \nn
\eea
The involution $\iota_{23}$ just corresponds to the interchange $(q_2,p_2) \leftrightarrow (q_3,p_3)$, so is clearly a canonical transformation.  The involution $\iota_1$ is also canonical, with generating function $S_1 = \frac{q_1P_1-q_2P_2+q_3P_3}{q_1^2-q_2^2-q_3^2}$.

\section{A Super-integrable System with Kinetic Energy (\ref{H12})}\label{concurv-super1}

In \cite{f17-5} we discuss the addition of (scalar) potentials to the kinetic energy (\ref{H12}), whilst retaining complete integrability, with an involutive system of 3 functions.  Since we only considered integrals which were at most {\em quadratic} in momenta, our completely integrable cases were inevitably separable and depended upon 3 arbitrary functions, each of a single variable (the separation coordinates).  For maximal super-integrability we required two additional integrals (also chosen to be quadratic in momenta), restricting the 3 arbitrary functions, which finally were fixed, up to 2 arbitrary parameters.  The 5 first integrals are
\bea
&& H = q_1^2(p_1^2-p_2^2-p_3^2)+ q_1^2\, \left(\frac{k_1}{q_2^2}+\frac{k_2}{q_3^2}\right), \quad  F_1 = e_1^2-\,\frac{k_1}{q_2^2}  =   p_2^2-\,\frac{k_1}{q_2^2},  \nn\\[-2mm]
   &&    \label{F14}    \\[-2mm]
&& F_2 = e_2^2 -\,\frac{k_2}{q_3^2} = p_3^2 -\,\frac{k_2}{q_3^2} ,\quad  F_3 = f_1^2-\,\frac{k_1 (q_2^2+q_3^2-q_1^2)^2}{q_2^2}  ,  \quad  F_4 = \frac{1}{4}f_2^2-\,\frac{k_2 (q_2^2+q_3^2-q_1^2)^2}{q_3^2},  \nn
\eea
where $f_1, f_2$ are defined in the list (\ref{diag-alg}).  In this case, we also find that $h_1$ is a first integral.

We therefore have $6$ first integrals $(H,F_1,F_2,F_3,F_4,h_1)$, but the rank of their Jacobian is $5$, so there should be an algebraic relation between them.  The involutions $\iota_1$ and $\iota_{23}$ act on these elements in a simple way, so they generate a Poisson algebra which also obeys such symmetry rules (see Table \ref{Tab:i123}).  Remarkably, we can use these symmetries to derive the entire table of Poisson relations, requiring a further 5 elements.

\begin{table}[h]\centering
\caption{The action of the involutions on the Poisson algebra and parameters}\label{Tab:i123}
$$
\begin{array}{|c||c|c|c|c|c|c|c|c|c|c|c|c|c|}\hline
& H & F_1 & F_2 & F_3 & F_4 & F_5 & F_6 & F_7 & F_8 & F_9 & h_1 & k_1 & k_2 \\\hline
\iota_1: & H & F_3 & F_4  & F_1 & F_2 & F_5 & F_6 & F_9 & F_8 & F_7 & -h_1 & k_1 & k_2\\\hline
\iota_{23}: & H & F_2 & F_1  & F_4 & F_3 & F_6 & F_5 & -F_7 & -F_8 & -F_9 & h_1 & k_2 & k_1\\ \hline
\end{array}
$$
\end{table}

We don't need to repeat the whole table of Poisson relations here (see \cite{f17-5}).  For our current purposes, the interesting integral is $h_1$, which is {\em linear} in momenta, so can be used in our Kaluza-Klein reduction. It satisfies the simple Poisson relations:
$$
 \{F_i,h_1\}=\lambda_i F_i,\;\; i=1,\dots ,9,\quad\mbox{where}\quad \lambda = (4,4,-4,-4,0,0,4,0,-4).
$$
For our reduction, we are interested in any functions of $F_i$ which commute with $h_1$.  First we have $F_5, F_6, F_8$, but there are also quadratic functions of $F_i$, such as $F_1F_3, F_1F_4,F_2F_3$ and so on, as well as higher order polynomials of $F_i$.  It follows from the Jacobi identity that this set of functions form a Poisson subalgebra and, when reduced, will continue to be first integrals of the reduced Hamiltonian.  There can be no more than 3 independent functions in the $2-$dimensional reduced space.  In this case the 3 functions $F_5, F_6, F_8$ form a subalgebra (with coefficients depending on $h_1$ and $H$):
{\small
\bea
\{F_5,F_6\} &=& 4 F_8,    \nn\\
\{F_5,F_8\} &=& H (4 F_6-h_1^2+4 H)+\frac{1}{2} F_5 (8 F_6-h_1^2+12 H)+2 F_5^2-2 (k_1+k_2) F_5 +\frac{1}{2} k_2 (h_1^2-4 H)-2 k_1 k_2,  \label{F568}  \\
\{F_6,F_8\} &=& -H (4 F_5-h_1^2+4 H)-\frac{1}{2} F_6 (8 F_5-h_1^2+12 H)-2 F_6^2+2 (k_1+k_2) F_6 -\frac{1}{2} k_1 (h_1^2-4 H)+2 k_1 k_2.  \nn
\eea   }
The formula for $\{F_6,F_8\}$ follows immediately from that of $\{F_5,F_8\}$ by an application of the involution $\iota_{23}$.
The functions $F_5, F_6$ are quadratic in momenta and can be written succinctly as
$$
F_5 = q_3^2F_1+q_1^2F_2+q_3p_3(q_3p_3+2 q_1p_1),\quad  F_6 = q_2^2F_2+q_1^2F_1+q_2p_2(q_2p_2+2 q_1p_1),
$$
with $F_8$ being the cubic result of the above Poisson bracket.  An explicit expression is given for it below, in the reduced coordinates.  The formula for $F_6$ follows from that of $F_5$ by an application of $\iota_{23}$.

\subsection{The Kaluza-Klein Reduction of (\ref{H12})}\label{kk-reduction1}

We now want to reduce our Hamiltonian (\ref{F14}) to the submanifold $h_1=b$.  To do this, we choose canonical coordinates $Q_i, P_i$ so that $h_1=P_3$, which makes $H$ independent of $Q_3$.  It is easy to see that
$$
\left\{\frac{q_1}{q_3},h_1\right\}=0,\quad \left\{\frac{q_2}{q_3},h_1\right\}=0,\quad \left\{-\frac{1}{2}\log q_3,h_1\right\}=1,
$$
so clearly we should choose $Q_3 = -\frac{1}{2}\log q_3$ and then some convenient functions of $z_1=\frac{q_1}{q_3},\, z_2=\frac{q_2}{q_3}$ for $Q_1, Q_2$.  To obtain a diagonal metric on our reduced space we need the orthogonality relation $\sum_{i,j=1}^3 g^{ij}\pa_iQ_1\pa_jQ_2=0$.  We choose $Q_1=\frac{q_1}{q_3}$ and $Q_2 = s\left(\frac{q_1}{q_3},\frac{q_2}{q_3}\right)$, leading to
$$
((z_1^2-1)\pa_{z_1}+z_1z_2\pa_{z_2})s(z_1,z_2)=0 \quad\Rightarrow\quad s(z_1,z_2) = \varphi\left(\frac{z_2^2}{z_1^2-1}\right).
$$
With
\be\label{Q123}
Q_1=\frac{q_1}{q_3},\quad Q_2 = \frac{\sqrt{q_1^2-q_3^2}}{q_2},\quad Q_3 = -\frac{1}{2}\log q_3,
\ee
we find
$$
H = \sum_{i,j=1}^3 \hat g^{ij}P_i P_j + \frac{k_1 Q_1^2Q_2^2}{Q_1^2-1} +k_2 Q_1^2,\quad\mbox{where}\quad
               \hat g^{ij} = \left(
                \begin{array}{ccc}
                  Q_1^2(1-Q_1^2) & 0 & -\frac{1}{2}Q_1^3 \\[1mm]
                  0 & \frac{Q_1^2Q_2^2(Q_2^2-1)}{1-Q_1^2} & \frac{Q_1^2Q_2}{2(1-Q_1^2)} \\[1mm]
                  -\frac{1}{2}Q_1^3 & \frac{Q_1^2Q_2}{2(1-Q_1^2)} & -\frac{1}{2}Q_1^2 \\
                \end{array}
              \right).
$$
\br[Curvature Properties]
The $3-$dimensional metric $\hat g^{ij}$ is constant curvature, with Ricci scalar $R=-6$.  The $2-$dimensional metric (whose inverse is the diagonal $2\times 2$ block of $\hat g^{ij}$) is also a constant curvature metric, with Ricci scalar $R=-2$.
\er
If we set $P_3=b$ and complete the square, to write this as a $2-$dimensional system of the form (\ref{kk-general1}), then we find the vector potential
$$
A_1 = \frac{b Q_1}{2(1-Q_1^2)},\;\; A_2 = \frac{b}{2Q_2(1-Q_2^2)}\quad\Rightarrow\quad  \pa_1 A_2-\pa_2 A_1 = 0,
$$
so we can make a gauge transformation to
\begin{subequations}
\bea
H &=& Q_1^2(1-Q_1^2) P_1^2 + \frac{Q_1^2Q_2^2(Q_2^2-1)}{1-Q_1^2}\, P_2^2 + \frac{k_1 Q_1^2Q_2^2}{Q_1^2-1} +k_2 Q_1^2 + \frac{b^2Q_1^2Q_2^2}{4(1-Q_1^2)(1-Q_2^2)}  \label{2dHam}  \\
  &=&  Q_1^2(1-Q_1^2) P_1^2+k_2 Q_1^2 + \frac{Q_1^2}{1-Q_1^2} \left(Q_2^2(Q_2^2-1)\, P_2^2 - k_1 Q_2^2 - \frac{b^2 Q_2^2}{4(Q_2^2-1)}\right) ,   \label{2dHam-sep}
\eea
\end{subequations}
where the second form shows that this Hamiltonian is separable in these coordinates.

\subsubsection{Isometries of the $2-$Dimensional Metric}

Since this $2-$dimensional metric is constant curvature, it must have 3 isometries, corresponding to Noether constants of the kinetic energy
\be\label{H0}
H_0 = Q_1^2(1-Q_1^2) P_1^2 + \frac{Q_1^2Q_2^2(Q_2^2-1)}{1-Q_1^2}\, P_2^2.
\ee
We could, of course, calculate these directly, but in the spirit of this paper we derive them as reductions from the $6-$dimensional algebra (\ref{diag-alg}).

Clearly $h_2$ commutes with both $h_1$ and $H_0$ of (\ref{H12}), so we write $h_2$ in the new coordinates
$$
h_2 = \frac{2 Q_1 \sqrt{Q_1^2-1}}{Q_2} \, P_1 -\frac{2 (Q_2^2-1)}{\sqrt{Q_1^2-1}}\, P_2 +\frac{b \sqrt{Q_1^2-1}}{Q_2},
$$
where the ``non-homogeneous'' term is a consequence of setting $P_3=b$ and is removed by the same gauge transformation we used above (or, equivalently, setting $b=0$).

There are also 2 {\em quadratic} expressions which commute with both $h_1$ and $H_0$ of (\ref{H12}), which, for $b=0$, are the squares of expressions which are {\em linear} in $P_1, P_2$:
$$
-e_1 f_1 = \left(Q_2 \sqrt{Q_1^2-1}\, P_2\right)^2 \quad\mbox{and}\quad -\frac{1}{2}e_2f_2 = \left(\frac{\sqrt{Q_2^2-1}\, \left(Q_1 (Q_1^2-1) P_1+Q_2 P_2\right)}{Q_2\sqrt{Q_1^2-1}}\right)^2.
$$
We therefore define
\begin{subequations}
\bea
&& K_1 = \frac{2 Q_1 \sqrt{Q_1^2-1}}{Q_2} \, P_1 -\frac{2 (Q_2^2-1)}{\sqrt{Q_1^2-1}}\, P_2,\qquad K_3 = 2 Q_2 \sqrt{Q_1^2-1}\, P_2,\nn\\[-2mm]
   &&    \label{2dkilling}   \\[-2mm]
   &&  K_2 = \frac{2Q_1\sqrt{(Q_1^2-1)(Q_2^2-1)}}{Q_2}\, P_1+ \frac{2\sqrt{Q_2^2-1}}{\sqrt{Q_1^2-1}}\, P_2,   \nn
\eea
which satisfy
\be\label{K123}
\{K_i,H_0\}=0,\quad \{K_1,K_2\}=2 K_3,\quad  \{K_2,K_3\}=2 K_1,\quad  \{K_3,K_1\}=2 K_2,
\ee
\end{subequations}
with Casimir $H_0 = -\frac{1}{4}\, (K_1^2+K_2^2+K_3^2)$.

\br
Notice that our definitions of $K_i$ give
$$
K_1^2=h_2^2,\;\; K_2^2 = -2 e_2 f_2,\;\; K_3^2=-4 e_1f_1,\quad\mbox{so}\quad K_1^2+K_2^2+K_3^2=-4\left(e_1f_1+\frac{1}{2}\, e_2f_2-\frac{1}{4}\, h_2^2\right),
$$
which is just a multiple of the Casimir (\ref{H03d}) without the $h_1$ term.
\er

\subsubsection{The Algebra $F_5,\, F_6,\, F_8$}

We can write the functions $F_5,\, F_6,\, F_8$ in the same coordinates and gauge without changing the Poisson relations (\ref{F568}).  The two quadratic functions are
\bea
F_5 &=&  -Q_1^2(1-Q_1^2) P_1^2-k_2 Q_1^2 - \frac{1}{1-Q_1^2} \left(Q_2^2(Q_2^2-1)\, P_2^2 - k_1 Q_2^2 - \frac{b^2 Q_2^2}{4(Q_2^2-1)}\right),\nn\\
F_6 &=& \frac{Q_1^2(Q_1^2-1)}{Q_2^2}\, P_1^2+\frac{(Q_2^2-1)(Q_1^2Q_2^2-1)}{Q_1^2-1}\, P_2^2-\frac{4 Q_1(Q_2^2-1)}{Q_2}\, P_1P_2  \nn\\
  &&    \hspace{6cm} -\frac{k_1Q_1^2Q_2^2}{Q_1^2-1}-\frac{k_2(Q_1^2-1)}{Q_2^2}-\frac{b^2}{4(Q_2^2-1)},\nn
\eea
showing that the quadratic part of $F_5$ is diagonal in these coordinates.  It is, in fact, the ``companion'' quadratic integral, associated with the separability in these coordinates.  The non-diagonal $F_6$ is the ``additional'' integral.

Just as the original functions (\ref{F14}) were written in terms of the isometry algebra (\ref{diag-alg}), we can write $F_5,\, F_6,\, F_8$ in terms of the symmetries (\ref{2dkilling}):
\bea
F_5 &=& \frac{1}{4} (K_1^2+K_2^2) -\frac{k_1Q_2^2}{Q_1^2-1} - k_2Q_1^2-\frac{b^2Q_2^2}{4(Q_1^2-1)(Q_2^2-1)} ,\nn\\
F_6 &=& \frac{1}{4} (K_1^2+K_3^2) -\frac{k_1Q_1^2Q_2^2}{Q_1^2-1} - \frac{k_2(Q_1^2-1)}{Q_2^2}-\frac{b^2}{4(Q_2^2-1)},\nn\\
F_8 &=& \frac{1}{8} K_1K_2K_3 +\left(\frac{b^2Q_1}{4(Q_2^2-1)}-k_1Q_1(Q_2^2-1)\right)P_1 \nn\\
&&   \hspace{5cm}  +\left(\frac{k_2(Q_1^2-1)(Q_2^2-1)}{Q_2}-\frac{k_1Q_2(Q_2^2-1)}{Q_1^2-1}-\frac{b^2Q_2}{4(Q_1^2-1)}\right)\, P_2.\nn
\eea
The leading order part of $F_8$ is easily obtained as a consequence of the Poisson relations (\ref{F568}) and (\ref{K123}).

\subsubsection{Simultaneous Diagonalisation of $H$ and $F_6$}

Each quadratic integral defines a $(2,0)$ Killing tensor of the metric with components $g^{ij}$.  We have seen that in coordinates $Q_i$, the two tensors $g^{ij}$ and $f_5^{ij}$ are simultaneously diagonalised, making these integrals the ``companion pair'' which correspond to separation of variables in those coordinates.

The existence of two quadratic integrals means that $H$ separates in a second coordinate system, now simultaneously diagonalising the two tensors $g^{ij}$ and $f_6^{ij}$.  Recall that our original system (\ref{F14}) was invariant under the action of a pair of involutions (\ref{i123}).  In particular $\iota_{23}$ corresponded to the simple canonical transformation, switching $q_2 \leftrightarrow q_3$.  Applying this to (\ref{Q123}), we get the alternative set of coordinates:
$$
\bar Q_1=\frac{q_1}{q_2},\quad \bar Q_2 = \frac{\sqrt{q_1^2-q_2^2}}{q_3},\quad Q_3 = -\frac{1}{2}\log q_2,
$$
which gives a transformation on our reduced $2-$dimensional space:
$$
\bar Q_1 = \frac{Q_1Q_2}{\sqrt{Q_1^2-1}},\qquad     \bar Q_2 = \frac{\sqrt{Q_1^2Q_2^2-Q_1^2+1}}{Q_2}.
$$
In these coordinates, $F_5$ and $F_6$ switch roles:
$$
\bar H=H(\bar Q,\bar P,k_2,k_1,b),\;\;  \bar F_5 =F_6(\bar Q,\bar P,k_2,k_1,b),  \;\;  \bar F_6 =F_5(\bar Q,\bar P,k_2,k_1,b),  \;\;   \bar F_8 = -F_8(\bar Q,\bar P,k_2,k_1,b),
$$
with $\bar F_6$ now being diagonalised.  The switching of parameters $k_1\leftrightarrow k_2$ is normally built into the definition of $\iota_{23}$ (see (\ref{i123})), but not in our change of coordinates.

\subsubsection{Conformal Coordinates}

In $2-$dimensions, every (pseudo-)Riemannian manifold is conformally flat, so we can find coordinates $(u,v)$, such that
$$
ds^2 = \frac{du^2\pm dv^2}{\varphi(u,v)},
$$
for some function $\varphi(u,v)$.  The corresponding kinetic energy takes the form $H_0 = \varphi(u,v) (p_u^2\pm p_v^2)$.  Starting with the separable Hamiltonian (\ref{2dHam-sep}), we know that the kinetic energy will transform to Liouville type, for which $\varphi(u,v)=\varphi_1(u)+\varphi_2(v)$.  To construct these coordinates (as functions of $(Q_1,Q_2)$), we use the Laplace-Beltrami operator.

For an $n-$dimensional (pseudo-)Riemannian space, with local coordinates $x^1,\dots ,x^n$ and metric $g_{ij}$, the Laplace-Beltrami operator is defined
by
\be   \label{lbf}
L_b f = \sum_{i,j=1}^n
             \frac{1}{\sqrt{g}} \frac{\pa}{\pa x^j}\left(  \sqrt{g} g^{ij}\frac{\pa f}{\pa x^i}\right),
\ee
where $g$ is the determinant of the matrix $g_{ij}$.  The coefficients of the second order derivative terms in the Laplace-Beltrami operator are
the coefficients of the {\em inverse metric} $g^{ij}$, which are just the coefficients in the kinetic energy in the Hamiltonian context.
Generally, the formula (\ref{lbf}) also has terms involving {\em first order} derivatives.

For the $2-$dimensional conformal case, the Laplace-Beltrami operator takes a very simple form: $L_bf=\varphi(u,v) (\pa_u^2\pm \pa_v^2)$, with {\em no first order terms}.  As a consequence, the coordinates $u$ and $v$ satisfy $L_bu=L_bv=0$.

The Laplace-Beltrami operator for the metric corresponding to (\ref{H0}) is
\be\label{lbQ12}
L_b = Q_1^2(1-Q_1^2) \pa_{Q_1}^2 + \frac{Q_1^2Q_2^2(Q_2^2-1)}{1-Q_1^2}\, \pa_{Q_2}^2-2 Q_1^3\, \pa_{Q_1} + \frac{Q_1^2Q_2(2Q_2^2-1)}{1-Q_1^2}\, \pa_{Q_2}.
\ee
We define new coordinates $u=w_1(Q_1), v=w_2(Q_2)$, which will still be separation variables for any nontrivial functions $w_i$.  To satisfy $L_bu=L_bv=0$, the functions $w_i$ must satisfy
$$
(Q_1^2-1) w_1''+2Q_1 w_1' =0,\quad Q_2(Q_2^2-1)w_2''+(2Q_2^2-1)w_2'=0,
$$
so we choose new coordinates
$$
u = \frac{1}{2}\, \log\left(\frac{1-Q_1}{1+Q_1}\right),\;\; v = \arctan \left(\frac{1}{\sqrt{Q_2^2-1}}\right) \quad\Rightarrow\quad  Q_1 = -\tanh u,\;\; Q_2 = \mbox{cosec}\, v.
$$
Our functions then take the form  {\small
\bea
H &=& \sinh^2 u \left(p_u^2+p_v^2-k_1\, \mbox{cosec}^2\, v+k_2\, \mbox{sech}^2\, u -\frac{b^2}{4} \sec^2 v\right),\nn\\
F_5 &=& - \sinh^2 u \, (p_u^2+k_2\, \mbox{sech}^2\, u) -\cosh^2 u\left(p_v^2-k_1 \, \mbox{cosec}^2 v-\frac{b^2}{4}\sec^2v\right),\nn\\
F_6 &=& - \sinh^2 u\, \sin^2 v\, p_u^2+\frac{1}{2}\sinh 2u\, \sin 2v\, p_up_v+(1-\cosh^2u\, \cos^2 v)\, p_v^2 \nn\\
   &&  \hspace{4cm}  +k_1 \sinh^2 u\, \mbox{cosec}^2 v + k_2 \, \mbox{sech}^2\, u\, \sin^2 v -\frac{b^2}{4} \tan^2 v, \nn\\
F_8 &=& \frac{1}{2}\, (-\sinh^2 u  \sin 2v\, p_u^2+  \sinh 2u \cos 2v\, p_u p_v+  \cosh^2 u\sin 2v\, p_v^2)p_v  \nn\\
  &&  +\frac{1}{2}\sinh 2u \left(\frac{b^2}{4} \tan^2 v -k_1 \cot^2 v\right) p_u  +\left(-k_1 \cosh^2 u \cot v+\frac{k_2}{2} \, \mbox{sech}^2\, u \sin 2v-\frac{b^2}{4} \cosh^2 u \tan v\right) p_v.  \nn
\eea  }

\section{A Second Super-integrable System with Kinetic Energy (\ref{H12})}\label{concurv-super2}

Again we start with the kinetic energy (\ref{H12}), but now consider a different potential, leading to a different Poisson algebra.
In \cite{f17-5} we derived the system with
\begin{subequations}\label{HF1234}
\be\label{F14-2}
H = q_1^2(p_1^2-p_2^2-p_3^2)+ h({\bf q}), \;\;  F_1 = \frac{1}{4} h_1^2+g_1({\bf q}), \;\; F_2 = \frac{1}{4}h_2^2+g_2({\bf q}),\;\; F_3 = e_1^2+g_3({\bf q}),\;\;  F_4 = e_2^2+g_4({\bf q}),
\ee
where $h_1, h_2, e_1, e_2$ are defined in the list (\ref{diag-alg}), and where
\bea
&& h=k_1q_1^2\, (q_1^2-q_2^2-q_3^2)+k_2 q_1^2-k_3\, \frac{q_1^2}{q_2^2}+k_4\, \frac{q_1^2}{q_3^2}, \quad g_1 = k_1\, (q_1^2-q_2^2-q_3^2)^2+k_2 (q_1^2-q_2^2-q_3^2)\nn\\[-2mm]
   &&   \label{hhee-pots}  \\[-2mm]
&& g_2 = k_3\, \frac{q_3^2}{q_2^2}-k_4\, \frac{q_2^2}{q_3^2}, \quad   g_3 = k_1 q_2^2 + \frac{k_3}{q_2^2},\quad  g_4 = k_1 q_3^2 - \frac{k_4}{q_3^2}, \nn
\eea
\end{subequations}
and considered the Poisson algebra generated by the 4 functions $F_i$.

\br
This system is not invariant under the action of the involution $\iota_1$ (see (\ref{i123})), which would take us to an equivalent one, involving $f_1, f_2$ instead of $e_1,e_2$.  However, it is invariant under $\iota_{23}$, under which $(H,F_1,F_2,F_3,F_4,k_1,k_2,k_3,k_4) \mapsto (H,F_1,F_2,F_4,F_3,k_1,k_2,-k_4,-k_3)$.
\er

For this paper, we consider the parametric reduction $k_3=k_4=0$, in which case $g_2=0$, so we can replace the quadratic integral $F_2$ by the geometric symmetry $h_2$.  We therefore need to select generators which Poisson commute with $h_2$.  We have $\{h_2,F_1\}=0$ and $\{h_2,F_3+F_4\}=0$, giving us 2 quadratic generators for our reduced Poisson algebra.

\subsection{Another Kaluza-Klein Reduction of (\ref{H12})}\label{kk-reduction2}

We now want to reduce our Hamiltonian (\ref{HF1234}) to the submanifold $h_2=b$.  This time, we note that
$$
\left\{q_1,h_2\right\}=0,\quad \left\{q_2^2+q_3^2,h_2\right\}=0,\quad \left\{\frac{1}{2}\,\arctan\left(\frac{q_3}{q_2}\right),h_2\right\}=1,
$$
and that $q_1$ and $q_2^2+q_3^2$ are orthogonal with respect to the metric of (\ref{H12}).  We therefore choose
\be\label{Q123-2}
Q_1=q_1,\quad Q_2 = \sqrt{q_2^2+q_3^2},\quad Q_3 = \frac{1}{2}\,\arctan\left(\frac{q_3}{q_2}\right),
\ee
and find
$$
H_0 = Q_1^2 (P_1^2-P_2^2)-\frac{Q_1^2}{4 Q_2^2}\, P_3^2.
$$
This $3-$dimensional metric has constant curvature, with Ricci scalar $R=-6$.  Since there are no electromagnetic terms, there is no need of a gauge transformation, so we can immediately reduce to the $2-$dimensional space with $P_3=b$. The $2-$dimensional metric is again constant curvature, with Ricci scalar $R=-2$.

Defining $I_1=F_1,\, I_2=F_3+F_4$, we find
\begin{subequations}
\bea
H &=&  Q_1^2(P_1^2+k_1 Q_1^2 + k_2) -Q_1^2 \left(P_2^2 + k_1 Q_2^2 + \frac{b^2}{4 Q_2^2}\right), \label{2dHam-2} \\
I_1 &=& (Q_1P_1+Q_2P_2)^2+k_1 (Q_1^2-Q_2^2)^2+k_2 (Q_1^2-Q_2^2),    \label{I1} \\
I_2 &=& P_2^2 + k_1 Q_2^2 + \frac{b^2}{4 Q_2^2}.   \label{I2}
\eea
\end{subequations}
Clearly, $H$ is separable in these coordinates, with $H$ and $I_2$ being the ``companion pair'' of quadratic integrals.  The Poisson bracket $\{I_1,I_2\}=4 I_3$ defines the cubic integral
\begin{subequations}
\be\label{cube2}
I_3 = (Q_1P_1+Q_2P_2) \left(P_2^2 + k_1 Q_2^2 + \frac{b^2}{4 Q_2^2}\right) -2 k_1 Q_1Q_2(Q_2P_1+Q_1P_2)-k_2 Q_2 P_2,
\ee
which satisfies the algebraic relation
\be\label{cube2-rel}
I_3^2 = I_1I_2^2-k_1(H-I_1)^2+k_2 (H-I_1)I_2-\frac{k_1b^2}{2}\, (H+I_1)+\frac{k_2b^2}{4} I_2-\frac{1}{16}\, b^2(k_1b^2+4k_2^2).
\ee
\end{subequations}
We can use this relation to derive the formulae for $\{I_1,I_3\}$ and $\{I_2,I_3\}$.  Since the left side is $2 I_3\{I_k,I_3\}$, and the right side has a factor $\{I_1,I_2\}$, we can cancel the $I_3$ factor to get polynomial expressions for these Poisson brackets:
$$
\{I_1,I_3\} = 4 I_1I_2+2 k_2 (H-I_1)+\frac{1}{2}\, k_2 b^2,\quad  \{I_2,I_3\} = -2 I_2^2-4 k_1 (H-I_1)+2 k_2 I_2+k_1 b^2.
$$

\subsection{Isometries of the $2-$Dimensional Metric}

Since this $2-$dimensional metric is constant curvature, it must have 3 isometries, corresponding to Noether constants of the kinetic energy
\be\label{H0-2}
H_0 = Q_1^2 (P_1^2 -P_2^2).
\ee
To derive the isometry algebra (in the spirit of this paper) we note that in the $6-$dimensional algebra (\ref{diag-alg}) we have
$$
\{h_2,h_1\}=0,\quad \{h_2,e_1^2+e_2^2\}=0,\quad  \{h_2,f_1^2+\frac{1}{4}\, f_2^2\}=0.
$$
Since $h_1$ is already an isometry of the $3-$dimensional metric, it naturally reduces to 2 dimensions.  The two {\em quadratic} expressions are again (when $b=0$) {\em squares} of linear expressions, so give us another two reduced isometries:
\begin{subequations}
\be\label{2dkilling-2}
K_1=-2(Q_1P_1+Q_2P_2),\quad K_2 = P_2,\quad K_3 = 2 Q_1Q_2 P_1+(Q_1^2+Q_2^2) P_2,
\ee
which satisfy the Poisson relations of the Lie algebra $\mathbf{sl}(2)$:
\be\label{Kpbs}
\{K_1,K_2\}=-2 K_2,\quad \{K_1,K_3\}= 2 K_3,\quad \{K_2,K_3\}=  K_1,
\ee
with (\ref{H0-2}) as the Casimir:
\be\label{cash0-2}
H_0 = \frac{1}{4}\, K_1^2-K_2 K_3.
\ee
\end{subequations}
Notice that
$$
K_2^2=e_1^2+e_2^2,\;\; K_3^2 = f_1^2+\frac{1}{4}f_2^2 \quad\Rightarrow\quad K_2^2K_3^2=\left(e_1f_1+ \frac{1}{2}\, e_2f_2\right)^2 + \left(e_2f_1 - \frac{1}{2}\, e_1f_2\right)^2,
$$
and that, from (\ref{cas2}), $\; e_2f_1 - \frac{1}{2}\, e_1f_2 = \frac{1}{2} b h_1$.  If we again set $b=0$, then this term vanishes, so (taking the negative root) we have $K_2 K_3 =-\left(e_1f_1+ \frac{1}{2}\, e_2f_2\right)$, so
$$
\frac{1}{4}\, K_1^2-K_2 K_3 = \frac{1}{4}\, h_1^2+e_1f_1+ \frac{1}{2}\, e_2f_2,
$$
which is again a reduction of the Casimir (\ref{H03d}).

The 3 integrals, $\{I_j\}_{j=1}^3$, can also be written in terms of $K_i$:
\bea
I_1 &=& \frac{1}{4}K_1^2+(Q_1^2-Q_2^2) (k_2+k_1(Q_1^2-Q_2^2)),\nn\\
I_2 &=& K_2^2 +k_2 Q_2^2+\frac{b^2}{4Q_2^2},\nn\\
I_3 &=& -\frac{1}{2}K_1K_2^2 +k_1 Q_2 ((Q_2^2-2Q_1^2)P_2-Q_1Q_2P_1)-k_2Q_2P_2+\frac{b^2(Q_1P_1+Q_2P_2)}{4 Q_2^2} .  \nn
\eea


\end{document}